\documentclass[prb,twocolumn,amssymb, amsmath, floatfix]{revtex4}
\usepackage{graphicx}

\newcommand{\cc}{\ensuremath{\textrm{c.c.}}}

\renewcommand{\vec}[1]{\ensuremath{\boldsymbol{\mathrm{#1}}}}
\newcommand{\grad}{\ensuremath{\nabla}}

\newcommand{\curl}{\ensuremath{\grad\!\times\!}}
\newcommand{\pd}[2]{\ensuremath{\frac{\partial #1}{\partial #2}}}

\newcommand{\smfrac}[2]{\ensuremath{{\textstyle\frac{#1}{#2}}}}

\begin{document}

\title{Suppression of Spontaneous Supercurrents in a Chiral p-Wave Superconductor}%

\author{Phillip E. C. Ashby}
\affiliation{Department of Physics and Astronomy, McMaster University, Hamilton, Ontario, Canada L8S 4M1}
\author{Catherine Kallin}
\affiliation{Department of Physics and Astronomy, McMaster University, Hamilton, Ontario, Canada L8S 4M1}
\date{\today}

\begin{abstract}
The superconducting state of SRO is widely believed to have chiral
p-wave order that breaks time reversal symmetry.  Such a state is expected
to have a spontaneous magnetization, both at sample edges and at
domain walls between regions of different chirality.  Indeed,
muon spin resonance experiments are interpreted as evidence of
spontaneous magnetization due to domain walls or defects in
the bulk. However, recent magnetic microscopy experiments place
upper limits on the magnetic fields at the sample edge and surface
which are as much as two orders of magnitude smaller than the
fields predicted theoretically for a somewhat idealized chiral
p-wave superconductor.  We investigate the effects on the spontaneous
supercurrents and magnetization of rough and pair breaking surfaces
for a range of parameters within a Ginzburg-Landau formalism.  The
effects of competing orders nucleated at the surface are also
considered.  We find the conditions under which the edge currents
are significantly reduced while leaving the bulk domain wall currents
intact, are quite limited.  The implications for interpreting the
existing body of experimental results on superconducting SRO within
a chiral p-wave model are discussed.
\end{abstract}

\maketitle

\section{Introduction}

Strontium ruthenate, Sr$_2$RuO$_4$ (SRO), has attracted considerable
experimental and theoretical study since its
discovery.\cite{Maeno:1994fk}  It was the first perovskite
superconductor to be discovered which did not contain copper 
and is believed to have unconventional pairing
symmetry.\cite{Mackenzie:2003lr,Murakawa:2004uq}  Numerous experimental results 
have been interpreted as evidence of a superconducting order parameter
with spin-triplet pairing\cite{Ishida:1998fk,Duffy:2000rt,Ishida:2001qy,Nelson:2004fj,Murakawa:2004uq} and 
broken time reversal symmetry.\cite{Luke:1998lr,Luke:2000yq,Kidwingira:2006kx,Xia:2006yq}  The simplest
order parameter consistent with these observations corresponds to 
a chiral p-wave, $p_x\pm ip_y$, Cooper pairing symmetry,\cite{Mackenzie:2003lr} 
analogous to the A phase of superfluid He-3.\cite{volovik88}

This state is expected to give rise to spontaneous supercurrents flowing 
along the sample edge,\cite{Matsumo:1999,Rice:2006fr,RahulRoy,Sigrist:1991lr} which are screened by the
Meissner effect so that the magnetic field is zero inside the
superconductor.  The net result is a magnetic field confined near the
edge of the sample.  These spontaneous currents and fields can also
occur within the sample at domain walls between $p_x+ip_y$ and $p_x-ip_y$ domains.\cite{Sigrist:1989lr,Sigrist:1991lr}
Surface currents and domain wall currents in a chiral p-wave superconductor
have been studied by Matsumoto and
Sigrist\cite{Matsumo:1999} and others\cite{volgork1985,Sigrist:1989lr,Sigrist:1991lr,Furusaki:2001yq,RahulRoy,Logoboy2008} and should be observable by scanning probe
measurements.\cite{Kwon:2003lr}  As well, muon spin resonance experiments have been interpreted as evidence for internal fields present at domain walls.\cite{Luke:1998lr,Luke:2000yq}

Recent scanning Hall bar and superconducting quantum interference
device (SQUID) microscopy measurements did not see the expected
signatures of spontaneous currents at the sample edges and 
surfaces.\cite{moler2005,Kirtley:2007kx}
These null measurements set upper limits on the spontaneous currents which
are approximately two orders of magnitude smaller than the values
predicted from simple chiral p-wave order.\cite{Kirtley:2007kx}
Given the considerable body of experimental results taken as evidence for
chiral p-wave order, it is important to understand whether the absence
of observable magnetization at the edges can be explained within a
theory of bulk chiral p-wave superconductivity.  One possibility
discussed by Kirtley {\it et al.}\cite{Kirtley:2007kx} is 
domains at the surface smaller than 1 or 2 microns on average. 
Given the size of the experimental probes, this could account for the
null results.\cite{Kirtley:2007kx}  
Indeed, Josephson tunneling measurements were interpreted
as evidence of chiral p-wave order with small dynamic domains,\cite{Kidwingira:2006kx}
although other results would be incompatible with such small domains at the 
surface\cite{Nelson:2004fj,Xia:2006yq} or in the bulk.\cite{Nelson:2004fj}  The formation of domain walls is energetically unfavorable in the Meissner state\cite{Sigrist:1991lr, Logoboy2008} and the samples are considered clean (otherwise T$_{\rm c}$ is noticeably reduced as expected for unconventional pairing\cite{Mackenzie:1998vn}), so such small domains arising from dynamics and pinning would be somewhat surprising.  However, an alternative to the Meissner state, one which favors domains of roughly the size of the penetration depth, has been proposed.\cite{Logoboy:2009kx}  Additional experiments are required to either rule out or confirm the presence of small domains.

Alternatively, one might expect surface roughness or other surface effects to reduce
the spontaneous currents and, in this paper, we investigate this possibility.  
Previously, only ideal (specular) surfaces of a chiral p-wave superconductor
have been considered,\cite{Matsumo:1999,Furusaki:2001yq} although
the effect of a rough surface has been considered for a neutral chiral p-wave
superfluid\cite{Nagato:1998ys} where screening currents are absent.  Rough surfaces
can be studied in the Bogoliubov-de Gennes (BdG) formalism or closely related Greens
function formalism.\cite{Nagato:1998ys}  Here, we use a Ginzburg-Landau (GL) formalism
allowing us to more readily study the effect
of a variety of surfaces as the parameters in the theory are varied.  These correspond to studying different microscopic Hamiltonians in the
BdG formalism, which each stabilize a $p_x\pm ip_y$ superconductor. 
The BdG formalism is more accurate at low temperatures, although for specular
surfaces it was found that the GL calculations gave qualitatively
similar results for the spontaneous currents and fields.\cite{Matsumo:1999,Furusaki:2001yq}  We also consider the effect of surfaces which nucleate a non-chiral p-wave
order parameter, while maintaining $p_x\pm ip_y$ in the bulk, as a possible
mechanism for suppressing the predicted edge currents. 

\section{Ginzburg-Landau Equations}

The Ginzburg-Landau free energy functional describing a single layer of Sr$_2$RuO$_4$ expressed in terms of dimensionless variables takes the form\cite{Sigrist:1991lr}

\begin{widetext}
  \begin{align}
    \label{eq:free}\nonumber F = \frac{\vec{H}_C^2\xi^3}{4\pi}\int d^3r &\left[-\smfrac{1}{2}\left(|u|^2 + |v|^2\right) + \left(\smfrac{1}{8} +\smfrac{1}{2}b_2\right)\left(|u|^2 + |v|^2\right)^2 +\smfrac{1}{2}b_2\left(u^*v - uv^*\right)^2 - \smfrac{1}{8}b_3\left(|u|^2-|v|^2\right)^2+k _1\left[|d_xu|^2 + |d_yv|^2\right]\right.\\
     &\left.+ k_2\left[|d_yu|^2 + |d_xv|^2\right] + k_3\left[(d_xu)^*(d_yv) +\cc \right]+k_4\left[(d_xv)^*(d_yu)+\cc \right] + \kappa^2\left(\curl{\vec{a}}\right)^2\right].
  \end{align}
  \end{widetext}
Here, $u$ and $v$ are the x and y-components of the order parameter, respectively.  The $b_i$ and $k_i$ are dimensionless material dependent constants, and $d_i = \pd{\hphantom{\psi}}{x_i} -ia_i$ are the usual gauge covariant derivatives.  The position $x$ is scaled by the coherence length, $\xi$, $\vec{a}$ is the dimensionless vector potential, and we have introduced $\kappa=\frac{\lambda}{\xi}$, the usual GL parameter.  Parameters satisfying $b_2>0$ and $b_3-4b_2<0$ stabilize the chiral p-wave state.  

The values of the coefficients in the free energy can be computed in the weak coupling limit of a BCS superconductor with triplet pairing aligned along $\vec{\hat{z}}$ as described by Furusaki {\it et al.}\cite{Furusaki:2001yq} and correspond to
$b_2=\frac{1}{8}$, $b_3=0$, $k_1=\frac{3}{4}$, $k_2 = \frac{1}{4}$,
$k_3=\frac{1}{4}$ and $k_4=\frac{1}{4}$. It is also sometimes convenient to introduce the variables $k_\pm = \frac{1}{2}(k_3\pm k_4)$. We take $\lambda = 190$
nm and $\xi = 66$ nm as parameters appropriate for strontium
ruthenate, unless noted otherwise.

The order parameters are parametrized by $u=|u|e^{i\theta}$ and
$v=|v|e^{i(\theta+\phi)}$.  We require the free energy to be
stationary with respect to variations of the order parameters and the
vector potential to obtain 6 coupled non-linear partial differential
equations. We consider the case of a boundary at $x=0$ with the superconductor occupying the half-plane $x>0$.  The symmetry in the problem allows us to discard y
derivatives, as well as to choose the gauge where $a_x = 0$.  Therefore, $u$ and $v$ are taken to be functions of $x$ only. This reduces the problem to the solution of the following 5 equations:

\begin{widetext}
\begin{align}
0 =& -k_1|u|^{''} -\left(k_++k_-\right)\left(a_y|v|\sin(\phi)\right)^{'} -\smfrac{1}{2}|u| +\left(\smfrac{1}{4}+b_2-\smfrac{1}{4}b_3\right)|u|^3 +\left(\smfrac{1}{4}+b_2\cos(2\phi)+\smfrac{1}{4}b_3\right)|v|^2|u| \nonumber\\
&+k_1|u|{\theta^{'}}^2+k_2|u|a_y^2 - k_+\cos(\phi)|v|a_y\left(2\theta^{'}+\phi^{'}\right) -k_+\sin(\phi)a_y|v|^{'}+k_-\cos(\phi)|v|a_y\phi^{'}+k_-\sin(\phi)a_y|v|^{'}\label{ueqn},\\
0 =& -k_2|v|^{''} +\left(k_+-k_-\right)\left(a_y|u|\sin(\phi)\right)^{'} -\smfrac{1}{2}|v| +\left(\smfrac{1}{4}+b_2-\smfrac{1}{4}b_3\right)|v|^3 +\left(\smfrac{1}{4}+b_2\cos(2\phi)+\smfrac{1}{4}b_3\right)|u|^2|v| \nonumber\\
&+ k_1|v|a_y^2 +k_2|v|{(\theta+\phi)^{'}}^2- k_+\cos(\phi)|u|a_y\left(2\theta^{'}+\phi^{'}\right)+k_+\sin(\phi)a_y|u|^{'}+k_-\cos(\phi)|u|a_y\phi^{'} +k_-\sin(\phi)a_y|u|^{'},\\
 0=&-k_2\left(|v|^2(\theta+\phi)^{'}\right)^{'} +\left(k_+ - k_-\right)\left(|u||v|a_y\cos(\phi)\right)^{'} - 2|u|^2|v|^2b_2\sin(\phi)\cos(\phi)+k_+|u||v|a_y\sin(\phi)\left(2\theta^{'}+\phi^{'}\right)\nonumber\\
 &+k_+a_y\cos(\phi)\left(|v||u|^{'} -|u||v|^{'}\right)-k_-|u||v|a_y\sin(\phi)\phi^{'}+k_-a_y\cos(\phi)\left(|v||u|^{'} +|u||v|^{'}\right),
\end{align}
\begin{align}
0=&-\kappa^2a_y^{''}+a_y\left(k_1|v|^2+k_2|u|^2\right)-k_+\cos(\phi)|u||v|\left(2\theta^{'} +\phi^{'}\right)+k_+\sin(\phi)\left(|v||u|^{'}-|u||v|^{'}\right)\nonumber\\
&+k_-cos(\phi)|u||v|\phi^{'} + k_-\sin(\phi)\left(|v||u|^{'}+|u||v|^{'}\right)\label{ayeqn},\\
0=&k_1\left(|u|^2\theta^{'}\right)^{'}+k_2\left(|v|^2(\theta^{'}+\phi^{'})\right)^{'}-2k_+(\cos(\phi)|u||v|a_y)^{'} .\label{theqn}
\end{align}
\end{widetext}

We integrate Eq. \ref{theqn} to obtain:
\begin{align}
\theta^{'} = \frac{2k_+\cos(\phi)|u||v|a_y-k_2|v|^2\phi^{'}}{k_1|u|^2+k_2|v|^2},
\end{align}
which is then used in Eqs. \ref{ueqn}-\ref{ayeqn} to reduce the problem to only 4 equations. These equations are solved self-consistently using  a numerical relaxation algorithm similar to that described by Thuneberg.\cite{Thuneberg:1987lr}  The equation for the current follows from Eqn. \ref{ayeqn} and the Maxwell equation $\curl \vec{B}=\frac{4\pi}{c}\vec{j}$. We identify the terms in the current proportional to the vector potential as the screening currents, and the others as the spontaneous surface currents. 

\section{Weak Coupling Results}
  
We consider a superconductor filing the half plane $x>0$ with a surface at $x=0$.  To derive the boundary conditions on the order parameters, we follow Ambegaokar {\it et al.}\cite{Ambegaokar:1974vn}  By considering quasiparticle trajectories, they show that a surface is always pair-breaking for the component of the order parameter which is normal to the surface.\cite{Ambegaokar:1974vn} This implies the boundary condition, $u(0)=0$.  Combining this with the restriction that no current should pass through the interface leads to the condition \begin{align}\left.\frac{|v|^{'}}{|v|}\right|_{x=0} = \mathrm{const.}\label{vbc}\end{align}
For a clean surface (specular scattering) next to an insulator the appropriate choice is $\mathrm{const.}=0$.\cite{deGennes}
  
\begin{figure}
  \includegraphics[angle=-90,width=3.5in,clip]{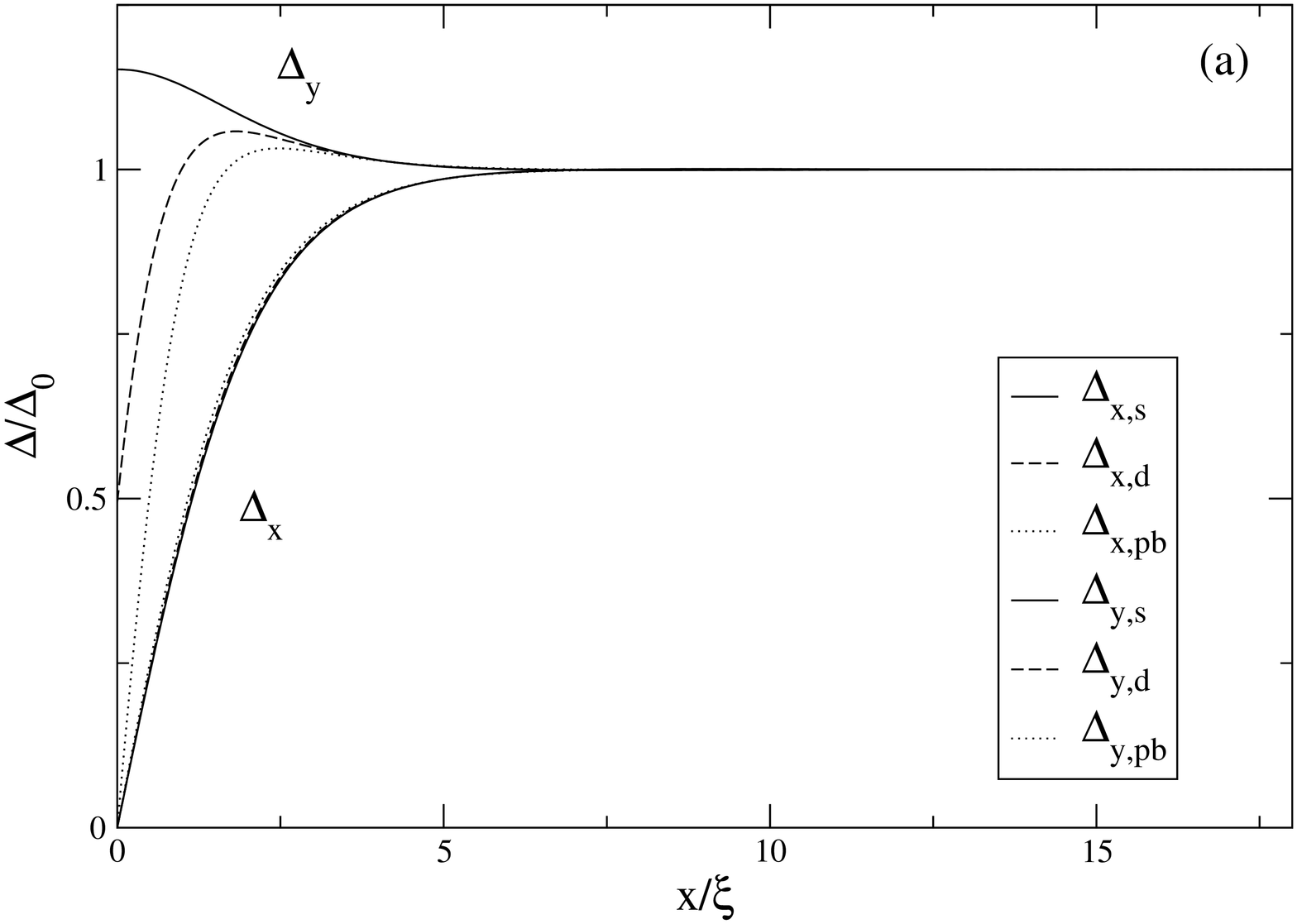}
  \includegraphics[angle=-90,width=3.5in,clip]{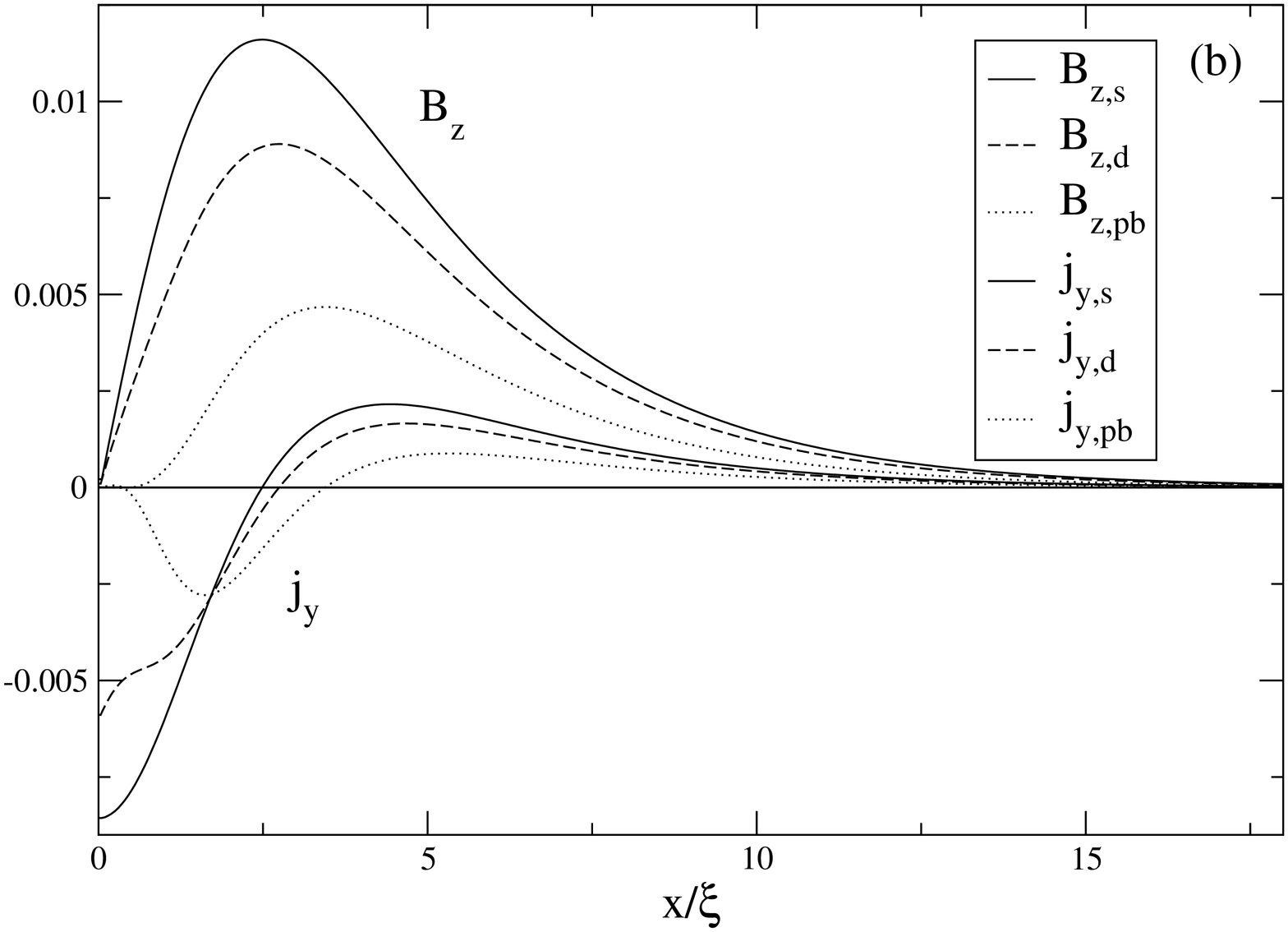}
  \caption{Self consistent solution of the GL equations
    for the weak coupling parameters.  (a)  The x and y components of
    the order parameters scaled by the bulk order parameter.  Here the
    subscripts s,d and pb denote the case of specular, diffuse, and pair breaking scattering, respectively.  (b) The magnetic field and current
    distributions scaled by $\frac{\hbar c}{2e\xi^2}$
    and $\frac{\hbar c^2}{8\pi e \xi^3}$ respectively.  A comparison
    of the integral of the magnetic field over 25 coherence lengths
    shows a 22\% reduction for the diffuse case compared to specular scattering.}
  \label{fig:weak}	
\end{figure}

The boundary condition on
$|u|$ suppresses it near the surface, and the term
$\left(\frac{1}{4}-b_2+\frac{1}{4}b_3\right)|u|^2|v|$ in the
equation of motion for $|v|$ allows $|v|$ to vary close to the surface.  In general,
if $(\frac{1}{2}-\alpha)\equiv\left(\frac{1}{4}-b_2+\frac{1}{4}b_3\right)>0$, $|v|$ will be
enhanced near the surface, and if
$\left(\frac{1}{2}-\alpha\right)<0$, it will be suppressed.\cite{Sigrist:1991lr}
For weak coupling, this coefficient takes the value $\frac{1}{8}$ and,
as seen in Fig. \ref{fig:weak}, the y component of the order parameter, $|v|$,
is larger at the surface than in the bulk.  

It is also noteworthy that even if we start from an
arbitrary relative phase, $\phi$, between the two order parameter components, self
consistent solution of the boundary problem forces the relative phase to $\pm
\frac{\pi}{2}$ even as the magnitude of the order parameters vary spatially at the surface. That
is,  the $p_x\pm ip_y$ state is maintained near the boundary.

An analysis of the GL equations for fixed $\phi=\frac{\pi}{2}$ shows that there are only two conditions under which the spontaneous current vanishes:\cite{ashbymsc} (i) $4b_2-b_3=0$ and (ii)  $k_1=k_2$, $k_3=k_4$, which gives $|v|=|u|$ everywhere.  The first case is a boundary of stability on the chiral p-wave state which requires $4b_2-b_3>0$. Thus, it follows that only with $|u|=|v|$ everywhere can the current be reduced to zero while maintaining chiral p-wave order in the bulk. Since by symmetry $|u|$ is suppressed at the surface, we must introduce an effect which also suppresses $|v|$ to reduce the current.  As a candidate, we examine the effects of a rough surface with variations on a scale much smaller than the coherence length, so that it can be treated as a boundary condition.  For such a surface the treatment of Ambegaokar {\it et. al} still applies\cite{Ambegaokar:1974vn}, $u(0)=0$, and $|v|$ satisfies Eqn. \ref{vbc}.

Following deGennes \cite{deGennes} the constant in Eqn. \ref{vbc} is denoted as $\frac{1}{b}$,
where $b=\infty$ corresponds to specular scattering and the limit of diffuse scattering provides a minimum value of $b=0.54$.\cite{Ambegaokar:1974vn}  We also consider an extra
suppression of the order parameter corresponding to $b<0.54$ which could
be caused by magnetic scattering at the surface which disrupts the
triplet pairing.  The limiting case $b=0$ corresponds to a completely pair breaking
surface with both components of the order parameter driven to zero at the surface.

The result of a self-consistent solution of the GL equations for the
weak coupling parameters is shown in Fig. \ref{fig:weak} for both
specular and diffuse scattering as well as for the fully pair breaking
boundary condition.  The x-component of the order
parameter is almost the same in all three cases since any surface along $\hat{\vec{y}}$ is fully pair breaking for this component.  As x approaches the surface, the y-component of
the order parameter still grows up as the x-component is
suppressed, but it also is ultimately suppressed close to the surface
due to the pair breaking boundary condition.  This behaviour can be attributed to the different healing lengths of the two components in response to a perturbation in $x$.  These qualitative shapes of the order parameters replicate those from previous work on the effect of a clean surface\cite{Matsumo:1999} and of a rough surface on a neutral chiral p-wave superfluid.\cite{Nagato:1998ys}  This demonstrates that the GL theory and boundary conditions treated here are in good agreement with the microscopic BdG and Green's function calculations.

\begin{figure}
  \includegraphics[angle=-90,width=3.5in,clip]{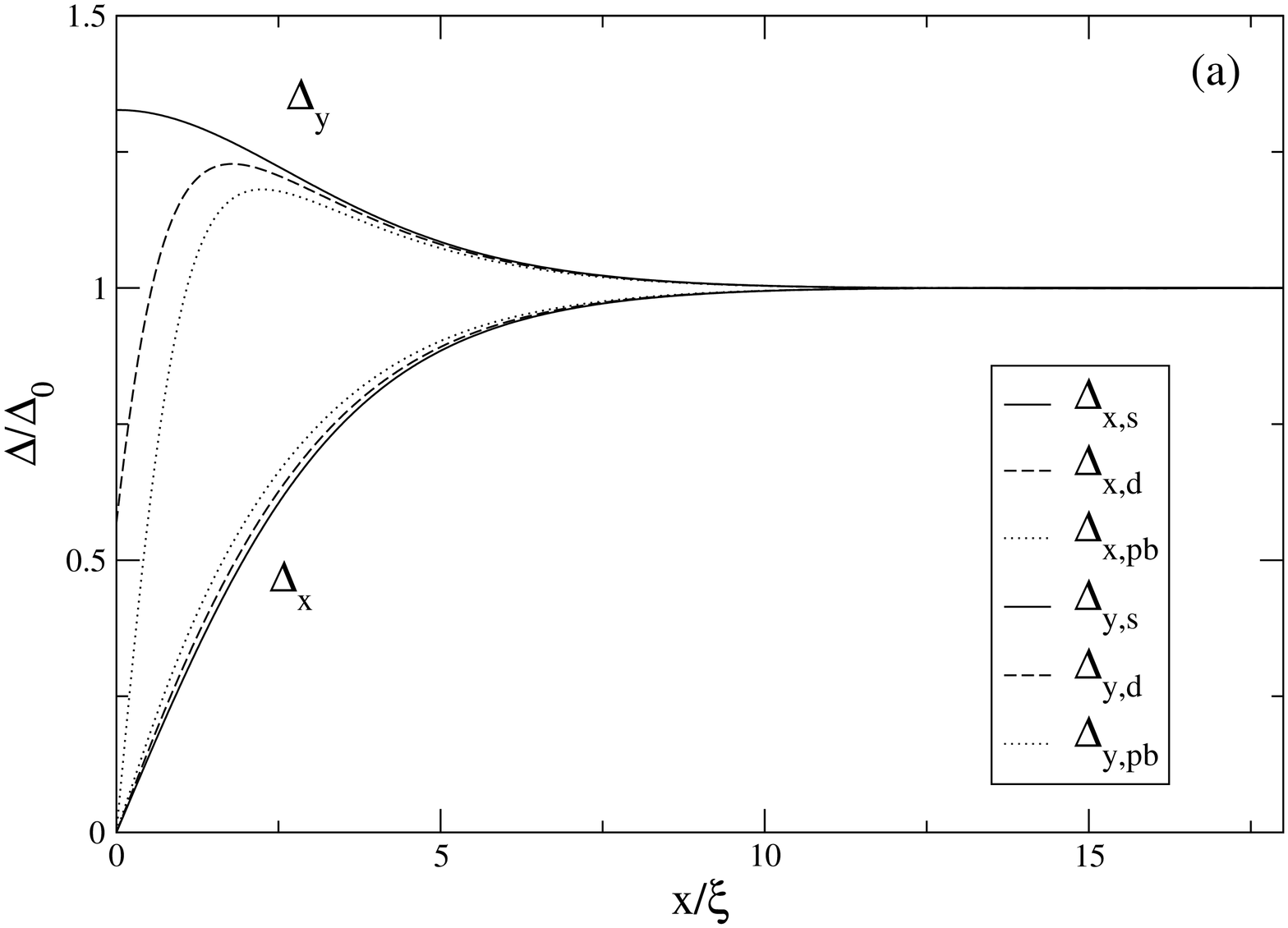}
  \includegraphics[angle=-90,width=3.5in,clip]{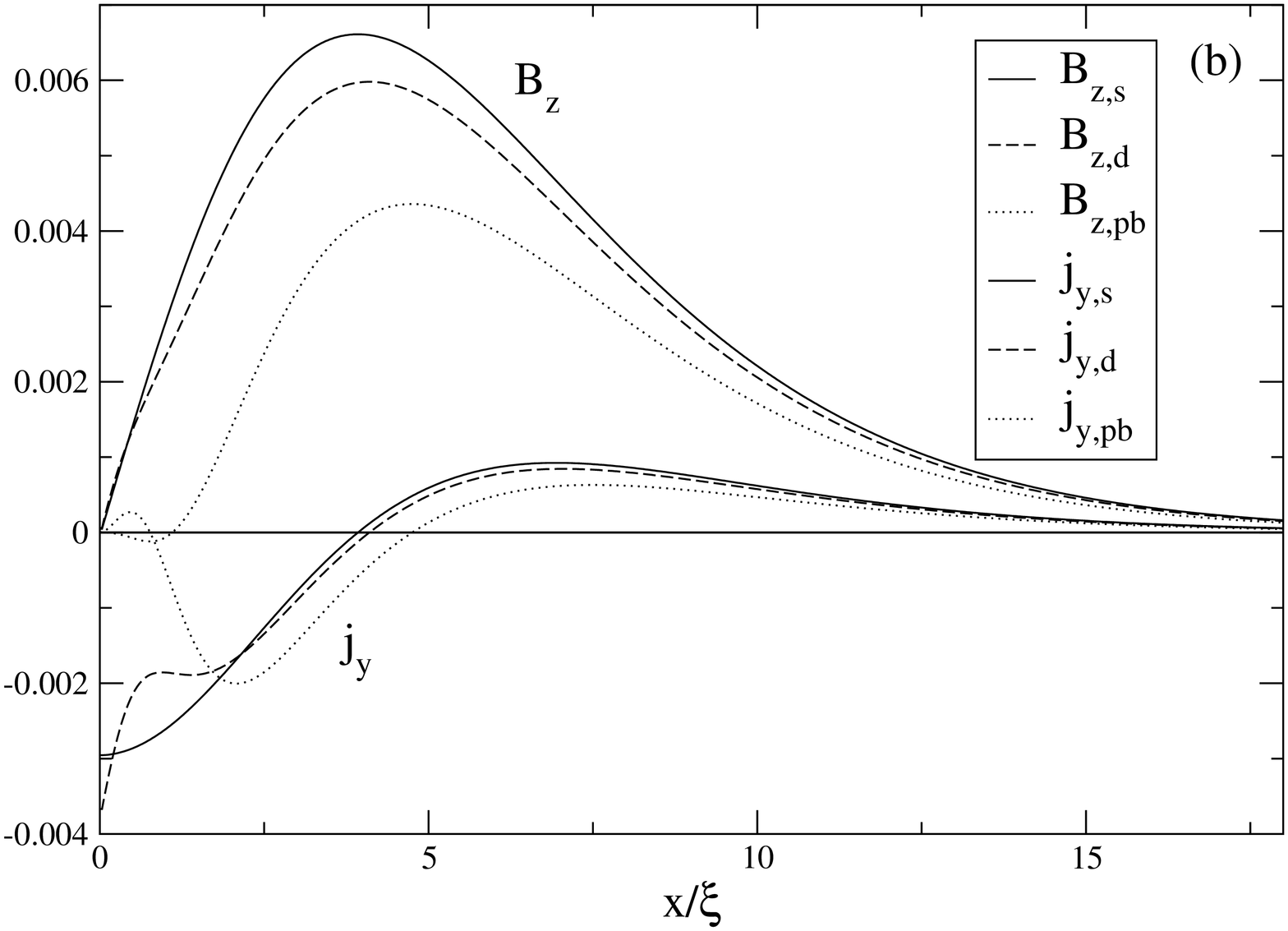}
  \caption{Self consistent currents, fields and corresponding order
    parameters for the parameters $b_2 = \frac{1}{16}$,
    $b_3=\frac{1}{8}$, $k_1=\frac{3}{4}$, $k_+=\frac{1}{4}$.  In this
    parameter regime the currents and fields are naturally suppressed,
    with the integrated magnetic field 23\% less than that of weak
    coupling parameters.  Also, the changes in the currents and fields
    due to surface roughness is reduced here resulting in only 10\%
    change in the integrated magnetic field.}
  \label{fig:bdiff}	
\end{figure}

\section{Results away from weak coupling}

Since the limit $4b_2-b_3\rightarrow0$ causes the currents to vanish, we examine self-consistently the dependence of the solutions on the parameter $4b_2-b_3$.  In Fig. \ref{fig:bdiff}  we demonstrate the dependance of
the fields and current for a value of the parameter $4b_2-b_3$ which
is closer to zero than for the weak coupling parameters.  We first notice that the healing length of the parameters is extended in this regime, resulting in a broader spontaneous current distribution near the edge. As this parameter is tuned closer to zero the spontaneous fields become
progressively smaller until the state is no longer stable and they
vanish.  However, the rough surface boundary condition has a smaller
effect on the reduction of the fields.  A reduction in magnetic field by changing parameters in this way will reduce all the magnetic signatures, and will not be able to account for the experiments taken as evidence for time reversal symmetry at domain walls.

\begin{figure}
  \includegraphics[angle=-90,width=3.5in,clip]{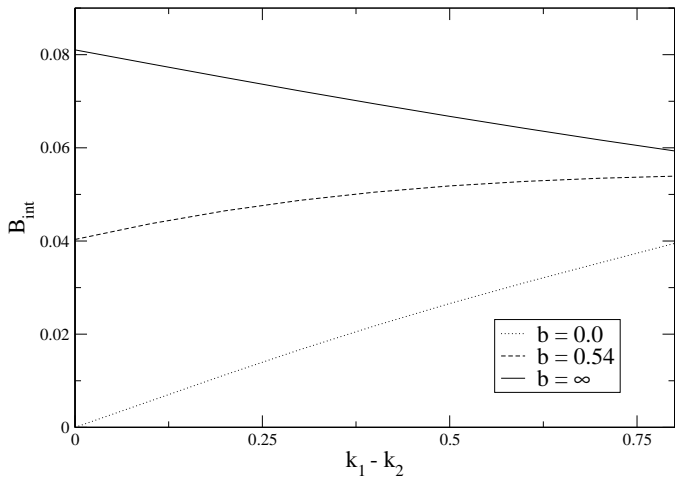}
  \caption{Dependence of the integrated magnetic field on the
    parameter $k_1-k_2$ for specular and diffuse scattering, and the
    pair breaking boundary condition, all other
    parameters are as for weak coupling.  The weak coupling
    parameters for $k_1$, $k_2$ correspond to $k_1-k_2 =
    \frac{1}{2}$.}
  \label{fig:kB}	
\end{figure}

Variation of the coefficients $k_i$, which are the stiffnesses of the
order parameters, also has an effect on the magnetic fields
produced at the surface . The weak coupling ratio
$\frac{k_1}{k_2}=3$ is apparent in both Figs. \ref{fig:weak} and
\ref{fig:bdiff}, as the x and y components heal over different
length scales. In Fig. \ref{fig:kB} we change the parameter $k_1$ and
observe the change in integrated magnetic field as the two order
parameters are forced to change on the same length scale.
Allowing $k_1$ and $k_2$ to become equal does not change the
integrated magnetic field significantly if the boundary conditions on $u$ and $v$ differ, as they do for specular or diffuse scattering.  On the other hand, if the surface is pair breaking and $v$ is significantly suppressed at the surface (in addition to $u$) then the integrated field falls off much faster. For $k_1=k_2$ the currents and fields are zero.  

It is interesting to ask if the parameter choice $k_1=k_2$ will also have a large effect on the currents produced at a domain wall.  There are two types of domain walls where:\cite{Sigrist:1991lr} (I) the relative phase continuously changes through the wall or (II) the phase changes discontinuously.  Recently a paper studied domain walls in the GL formalism and found a parameterization which connected the two types of domain walls.\cite{Logoboy2008}  They expressed the magnetization as a function of the GL parameters.  Previously, Matsumoto and Sigrist showed that the domain wall configuration of type II is energetically favoured.\cite{Matsumo:1999} Since only one order parameter is driven to zero for this type of domain wall, our previous analysis shows that a non-zero magnetization will be produced, consistent with the result of Logoboy and Sonin.\cite{Logoboy2008} This situation would allow for the magnetic signals attributed to domains in the bulk, as well as for the lack of currents at the edge.  The difference in magnitude of $k_1$ and $k_2$ is associated with the different energy costs of longitudinal and transverse perturbations respectively. There there is no symmetry which would require that $k_1=k_2$. In general one would expect the longitudinal fluctuations to be stiffer, as is the case for the weak coupling parameters. 

One last feature of Fig. \ref{fig:kB} which stands out, is the increase of the overall magnetic signal for the specular boundary condition as $k_1\rightarrow k_2$.  To understand this we express the screening currents ($j_s$) in terms of the sum and difference of $k_1$ and $k_2$:
\begin{align}
\label{screenhole}\kappa^2j_s&=-a_y(k_1|v|^2+k_2|u|^2)\nonumber\\
&=-\frac{a_y}{2}\left[(|u|^2+|v|^2)+(k_1-k_2)(|v|^2-|u|^2)\right],
\end{align}
where we have used the constraint $k_1+k_2=1$.
From Eq. \ref{screenhole} we see that as $k_1-k_2\rightarrow0$ the screening currents are reduced in the region where $(|v|^2-|u|^2)>0$, which is only satisfied near the sample edge.  The change in $k_1-k_2$ also changes the spatial dependence of the spontaneous currents.  The spontaneous current both increases in magnitude and is pulled closer to the sample edge as $k_1$ is reduced.  This move of the spontaneous currents to the region where the screening currents are reduced results in an increased magnetic signal.

\begin{figure}
  \includegraphics[angle=-90,width=3.5in,clip]{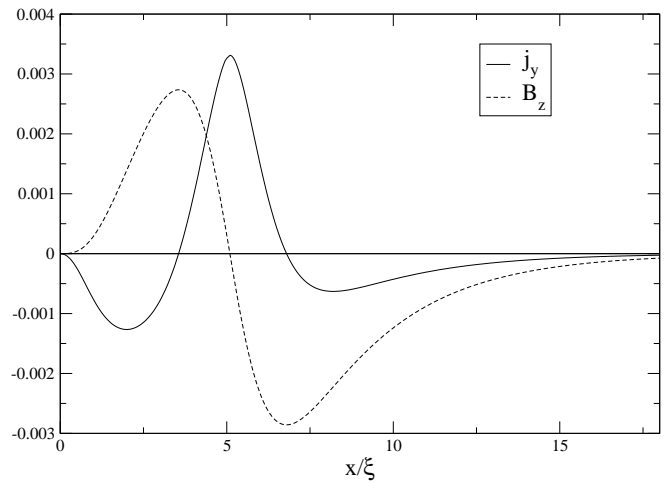}
  \caption{Computed currents and fields for a solution with spatially varying
    relative phase near the surface.  The parameters used are those for weak coupling, with
    $b_2$ negative for $x\leq 5$ but positive elsewhere.}
  \label{fig:p+p}	
\end{figure}

\section{Effect of competing surface order}

To consider the effect on the edge currents of competing order which is favoured by
the surface, we consider several possibilities.  First, we allow the parameters that stabilize the chiral p-wave state to vary spatially near the surface. In particular we change the sign of the $b_2$ term in the free energy in a region near the surface ($b_2>0$ favours the $p_x+ ip_y$
state, and $b_2<0$ the $p_x+p_y$ state).  The $p_x+p_y$ ground state does not support spontaneously generated supercurrents, and being stabilized near the edge could reduce magnetic signatures. The resulting currents and fields from a self-consistent calculation are shown in
Fig. \ref{fig:p+p}. Notice that both the overall magnitude of the
fields is suppressed, and there is a change in sign.  This
alternating magnetic field is much like at a domain wall and will also
affect the magnitude of the measured magnetic fields as described by
Kirtley {\it et al.}\cite{Kirtley:2007kx}.  While these spatially alternating magnetic fields could reduce the measured signals, this scenario has a different order at the surface and is incompatible with the interpretation of some of the surface tunneling measurements.\cite{Nelson:2004fj,Kidwingira:2006kx}

We also consider the effect of subdominant non-chiral p-wave order coexisting with the chiral p-wave state.  This can be modeled by adding the following terms to the free energy:

\begin{align}
  f_2 &=a_2|w|^2 + b_4|w|^4+k_5\left(\left|D_xw\right|^2+\left|D_yw\right|^2\right)\nonumber\\
  &+b_5|w|^2\left(|u|^2+|v|^2\right)+b_6\left[w^{*2}\left(u^2+v^2\right)+\cc\right].
\end{align}
These terms would be caused by the addition of any one of the other
unitary states allowed under the crystal symmetry.  However, symmetry does not allow gradient terms coupling the new order
parameters to the old ones.  This means that even when the parameters
are such that this new order parameter can grow up near the surface,
it has little effect on the shape of the old order parameters and
hence, causes insignificant changes to the currents and fields, unless the transition temperature for this competing order (determined by $a_2) $is high enough.

Since we wish to maintain chiral p-wave order in the bulk, we consider an $a_2$ which varies spatially.  In particular we consider the case where the $T_c$ of $w$ is greater than that of the chiral p-wave near the surface.  Self consistent solutions of the GL equations show that the new order
parameter grows up  at the surface, suppressing
the chiral p-wave state which is recovered in the bulk.  This
configuration gives rise to an
alternating magnetic field, of similar magnitude to that shown
in Fig. \ref{fig:p+p}, with the maximum magnitude of the fields being about 30\% smaller.

\section{Discussion}

In an attempt to reconcile the results of various experiments on strontium ruthenate, we have explored a variety of mechanisms that could be responsible for
reducing the spontaneous edge supercurrents generated by a $p_x+ip_y$
superconductor, while simultaneously maintaining the $p_x+ip_y$ state in the bulk.  One can think of several possibilities for reducing all spontaneous currents, at edges and at domain walls, such as multiband effects or using GL parameters near the boundary of stability for $p_x+ip_y$.  However, one would then need to look for alternative explanations for the $\mu$SR measurements which have been taken as evidence for chiral p-wave domain walls.  Therefore we have focused on effects  which reduce the edge currents, but not the domain wall currents.

In particular, we examined the effect on the spontaneous supercurrents of rough and pair breaking surfaces, as well as the dependence on the GL parameters.  The spontaneous edge currents are zero or vanishingly small only in a very small region of parameter space in the presence of a fully pair breaking surface.  This tuning of parameters is unlikely to be realized in a physical system as it requires the coefficients for longitudinal and transverse gradients to be equal.

The effect of nucleating a non-chiral order parameter at the surface, while maintaining chiral p-wave order in the bulk, was also investigated.  This can give rise to solutions where the magnetic field alternates in sign near the surface.  These alternating magnetic signatures could produce null results for the edge currents if the length scale of the alternating magnetic field was sufficiently short.  Again, this scenario would be difficult to reconcile with the tunneling measurements.\cite{Nelson:2004fj,Kidwingira:2006kx}

Leggett has proposed an alternative wavefunction that reduces to the BCS wavefunction for the case of s-wave pairing\cite{leggett}.  While, for the chiral p-wave case, the BCS wavefunction predicts an angular momentum of the condensate of Cooper pairs given by $\frac{N\hbar}{2}$,\cite{Mermin:1980zr,RahulRoy} Leggett finds that the angular momentum of the condensate described by his wavefunction is $\frac{N\hbar}{2}\left(\frac{\Delta}{\epsilon_f}\right)^2$.  This large suppression of the angular momentum would indeed reduce spontaneous surface currents. However, this would also result in a suppression of all magnetic signatures and thus would leave the positive $\mu$SR results unexplained.

In summary, we have identified a number of possible explanations for the absence of observable edge currents in a chiral p-wave superconductor.  However, each of these is then inconsistent with the interpretation of tunneling and/or $\mu$SR results that are interpreted as a direct observation of the fields induced by supercurrents at domain walls. It would be illuminating to investigate this in greater detail by modeling the $\mu$SR lineshapes expected for chiral p-wave domain walls.
\begin{acknowledgments}
  The authors would like to thank M.~Sigrist,  A.~J.~Berlinsky, and R.~Roy for useful discussions.  This work was supported by NSERC and the Canadian Institute for Advanced Research.
  \end{acknowledgments}
\bibliography{biblio}

\end{document}